\documentclass[aps,twocolumn,superscriptaddress]{revtex4-1}
\usepackage{}
\usepackage{amssymb}
\usepackage{amsfonts}
\usepackage{graphics,graphicx,epsfig,bm,amsmath,amsthm,amssymb}
\usepackage{bm}
\usepackage{bbm}
\usepackage{longtable}
\usepackage{multirow}
\usepackage{array}
\usepackage{color}
\usepackage[usenames,dvipsnames]{xcolor}

\usepackage{float}
\usepackage[a4paper,colorlinks=true,
linkcolor=blue,citecolor=blue,
pdfauthor={ },
pdftitle={ },
pdfsubject={ },
pdfkeywords={ }]{hyperref}

\begin{document}

\title{Observation of dynamical phase transitions in a topological nanomechanical system}

\author{Tian Tian}
\affiliation{CAS Key Laboratory of Microscale Magnetic Resonance and Department of Modern Physics, University of Science and Technology of China, Hefei 230026, China}
\affiliation{Synergetic Innovation Center of Quantum Information and Quantum Physics, University of Science and Technology of China, Hefei 230026, China}

\author{Yongguan Ke}
\affiliation{Laboratory of Quantum Engineering and Quantum Metrology, School of Physics and Astronomy, Sun Yat-Sen University (Zhuhai Campus), Zhuhai 519082, China}
\affiliation{State Key Laboratory of Optoelectronic Materials and Technologies, Sun Yat-Sen University (Guangzhou Campus), Guangzhou 510275, China}

\author{Liang Zhang}
\affiliation{CAS Key Laboratory of Microscale Magnetic Resonance and Department of Modern Physics, University of Science and Technology of China, Hefei 230026, China}
\affiliation{Synergetic Innovation Center of Quantum Information and Quantum Physics, University of Science and Technology of China, Hefei 230026, China}

\author{Shaochun Lin}
\affiliation{CAS Key Laboratory of Microscale Magnetic Resonance and Department of Modern Physics, University of Science and Technology of China, Hefei 230026, China}
\affiliation{Synergetic Innovation Center of Quantum Information and Quantum Physics, University of Science and Technology of China, Hefei 230026, China}

\author{Zhifu Shi}
\affiliation{CAS Key Laboratory of Microscale Magnetic Resonance and Department of Modern Physics, University of Science and Technology of China, Hefei 230026, China}
\affiliation{Synergetic Innovation Center of Quantum Information and Quantum Physics, University of Science and Technology of China, Hefei 230026, China}
\affiliation{Hefei National Laboratory for Physical Sciences at the Microscale, University of Science and Technology of China, Hefei 230026, China}

\author{Pu Huang}
\affiliation{National Laboratory of Solid State Microstructures and Department of Physics, Nanjing University, Nanjing 210093, China}

\author{Chaohong Lee}
\email{lichaoh2@mail.sysu.edu.cn}
\affiliation{Laboratory of Quantum Engineering and Quantum Metrology, School of Physics and Astronomy, Sun Yat-Sen University (Zhuhai Campus), Zhuhai 519082, China}
\affiliation{State Key Laboratory of Optoelectronic Materials and Technologies, Sun Yat-Sen University (Guangzhou Campus), Guangzhou 510275, China}

\author{Jiangfeng Du}
\email{djf@ustc.edu.cn}
\affiliation{CAS Key Laboratory of Microscale Magnetic Resonance and Department of Modern Physics, University of Science and Technology of China, Hefei 230026, China}
\affiliation{Synergetic Innovation Center of Quantum Information and Quantum Physics, University of Science and Technology of China, Hefei 230026, China}
\affiliation{Hefei National Laboratory for Physical Sciences at the Microscale, University of Science and Technology of China, Hefei 230026, China}


\date{\today}

\begin{abstract}
  Dynamical phase transitions (DPTs), characterized by non-analytic behaviors in time domain, extend the equilibrium phase transitions to far-from-equilibrium situations. It has been predicted that DPTs can be precisely identified by the discontinuities of the Pancharatnam geometric phase (PGP) during the time evolution. However, PGP always mixes with dynamical phase and the experimental observation of DPTs by PGP is still absent. Here, we theoretically present a novel scheme for eliminating the dynamical phase by taking advantage of chiral symmetry in the Su-Schrieffer-Heeger (SSH) model, and experimentally observe DPTs by directly measuring PGP in a quenched topological nanomechanical lattice. Time-dependent topological structures of the SSH model are configured by eight strong-coupled high-quality-factor nanomechanical oscillators. By measuring the vibration phase and the normalized amplitude of the edge oscillator, we show a direct classical analog of DPTs. Furthermore, we experimentally demonstrate the robustness of DPTs against weak structure disorders, and numerically explore the relation between DPTs and the equilibrium phase boundary. This work not only establishes the quantitative method to identify DPTs, but also opens the door for studying non-equilibrium topological dynamics with a well-controlled nanomechanical system.
\end{abstract}

\maketitle

\section{Introduction}
In recent years, non-equilibrium phenomena beyond the Ginzburg-Landau paradigm have attracted tremendous attentions and interests~\cite{Heyl2013,Heyl2018,Budich2016,Polkovnikov2011,Eisert2015,Schreiber2015,Smith2016,Choi2016,ZhangJ2017,Choi2017,Jurcevic2017,Zhang2017,Bernien2017,Flaschner2018,Wang2017,Qiu2018,Zhang2018,Vajna2015,Schmitt2015,Sun2018,ZHuang2016,Heyl2017,Bhattacharya2017}. In analog to equilibrium phase transitions, dynamical phase transitions (DPTs) are a kind of outstanding non-equilibrium phenomena which manifest non-analytic behaviors of the rate function, i.e. the logarithm function of returning probability to initial state in the time domain~\cite{Heyl2013,Heyl2018}. The DPT process is a global quench of the initial Hamiltonian's eigenstate under the finial Hamiltonian~\cite{Heyl2018}. So far, DPTs in sudden quenches have been found by measuring the rate function or observing the dynamical vortices~\cite{Jurcevic2017,Zhang2017,Bernien2017,Flaschner2018}. Theoretically, in a DPT, the non-analytic behavior occurs and the Pancharatnam geometric phase (PGP)~\cite{Pancharatnam1956,Samel1988} jumps $\pi$ simultaneously at the critical time when the returning probability goes through zero~\cite{Budich2016}. This means that the PGP provides an exact quantitative hallmark of DPTs. To the best of our knowledge, as the PGP in a quenched system always accompanies and mixes with the dynamical phase~\cite{Heyl2018,Budich2016,Flaschner2018,Qiu2018}, there is still no experiment observation of DPTs by direct measuring PGP.

On the other hand, sudden quench dynamics have been extended to topological systems~\cite{Heyl2018,Budich2016,Vajna2015,ZHuang2016,Schmitt2015,Flaschner2018,Wang2017,Qiu2018,Zhang2018,Sun2018,Heyl2017,Bhattacharya2017}.
Considering sudden quench from an initial bulk state of a trivial Hamiltonian to a nontrivial Hamiltonian, the equilibrium topology of the final Hamiltonian can be deduced by measuring linking number~\cite{Wang2017,Tarnowski2017}, or emerging ring structure~\cite{Zhang2018,Sun2018}.
Many researches in DPTs explore the connection to equilibrium topological phase transitions~\cite{Vajna2015,ZHuang2016,Schmitt2015,Flaschner2018,Wang2017,Qiu2018}.
Several theoretical works show that a topology-changing quench is guaranteed for the DPTs~\cite{Vajna2015,ZHuang2016,Sedlmayr2018}.
However, all these studies focus on quenches from a bulk state in a trivial phase~\cite{Vajna2015,Schmitt2015,Budich2016,ZHuang2016,Flaschner2018,Wang2017,Qiu2018,Zhang2018,Sun2018}. As a prominent signature of topology phases~\cite{Wang2009,Rechtsman2013,Susstrunk2015,He2016,St-Jean2017}, quenches from edge states are still an open question in studying DPTs.

Here, we present a novel scheme to directly measure PGP by eliminating the dynamical phase in sudden quenches from a topological edge state and explore DPTs in a reconfigurable nanomechanical lattice. In theory, considering the Su-Schrieffer-Heeger (SSH) model - a typical topological model~\cite{Su1979}, we choose a topological edge state as the initial state according to our quench protocol. Due to its chiral symmetry~\cite{Asboth2016}, the initial edge state equally populates all the symmetrical pairs of the final eigenstates so that the dynamical phase is naturally eliminated, regardless of the quenching being to a trivial Hamiltonian or a nontrivial one. This makes it possible to quantitatively reveal DPTs by directly measuring PGP. In experiment, we fabricate high-quality-factor nanomechanical lattice of eight oscillators, whose couplings are fully controllable. The quality factor of the oscillators can reach about $1 \times10^5$ at 77 K in vacuum, which is far better than the previous mechanical lattice~\cite{Huang2016} and this ensures the observation of dynamical evolutions. Different topological structures can be flexibly engineered by capacitive couplings~\cite{Rugar1991,Huang2013} between oscillators. The initial excitation and the high-resolution measurement are enabled with the standard magneto-motive technique~\cite{Cleland1999}. The motions of the parametrically coupled nanomechanical oscillators can be mapped onto the quantum analog of the tight-binding model, therefore we give the direct analog of DPTs in classical coupled oscillators. Since the dynamical phase is naturally eliminated, we directly observe the PGP and its jumps at critical times by demodulating the motion of the edge oscillator. The non-analytic behaviors of the rate function can also be observed by the normalized amplitude of the edge oscillator, which provides an original signature of the DPTs. This accomplishes the exploration of the relation between PGP and DPTs in experiment. Afterwards we introduce structure imperfections in the quench, and then the experimental result demonstrates the robustness of the topologically protected DPTs via measuring PGP.

The relation between DPTs and equilibrium phase boundary is studied in this article. We numerically calculate the quenches from different initial edge states and obtain the diagram which is determined by the presence or absence of DPTs. We find that DPTs always appear when quenches across the underlying equilibrium phase boundary, and there is strong connection between robust DPTs and the underlying topological phases.

\begin{figure}
  \includegraphics[width=1\columnwidth]{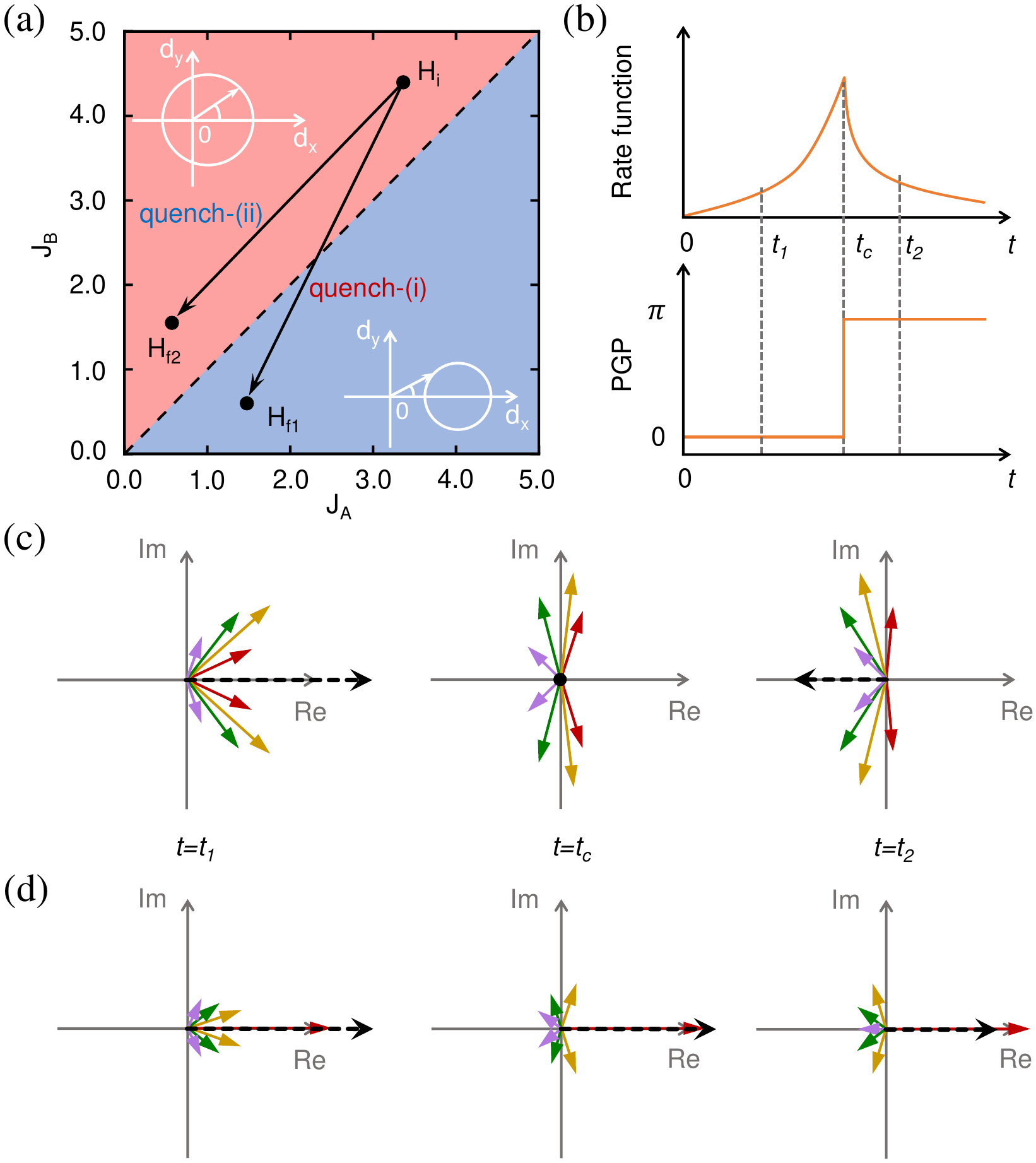}
  \caption{
  The schematic of quench dynamics from a topological edge state.
  (a) Equilibrium phase diagram of SSH model. If intracell hopping $J_{A}$ is larger than intercell hopping $J_{B}$, SSH chain is trivial and the winding number $\mathcal W =0$. Otherwise, the SSH chain is topological and the winding number $\mathcal W =1$. Two sudden quenches in a SSH model with an initial edge state are studied in this work: (i) from the topological nontrivial $\hat H_{i} (J_A< J_B)$ to trivial $\hat H_{f1} (J_A>J_B)$, and (ii) from the topological nontrivial $\hat H_{i} (J_A< J_B)$ to nontrivial $\hat H_{f2}$ ($J_A<J_B$). (b) Rate function and PGP at DPTs. In DPT, the non-analytic behavior of rate function occurs and the PGP simultaneously jumps $\pi$ at the critical time $t_{c}$. (c),(d) The Loschmidt amplitudes in quench-(i) and quench-(ii) at different times, respectively. All of the partial Loschmidt amplitudes showed by colored arrows always have symmetrical pairs because of chiral symmetry. For quench-(i) in (c), the magnitude and direction of the Loschmidt amplitude (dashed-black arrow) are changed with time. At the critical time $t_{c}$, the Loschmidt amplitude across zero, there is a DPT and PGP jumps $\pi$, see orange line in (b). For quench-(ii) in (d), the Loschmidt amplitude is dominated by zero mode (red arrow), and only the magnitude is changed with in this case. There is no DPT and PGP remains zero.}
  \label{schematic}
\end{figure}

\section{Quench dynamics in SSH model}

In this section, different from the quench scheme in the momentum space of SSH model~\cite{Vajna2015}, we put forward a theory of the quenches from a topological edge state for open boundary condition in real space. We present the scheme of naturally eliminating the dynamical phase and provide the theoretical pictures of DPTs in terms of Loschmidt vectors.

\subsection{Quench scheme}
\begin{figure}
	\includegraphics[width=1\columnwidth]{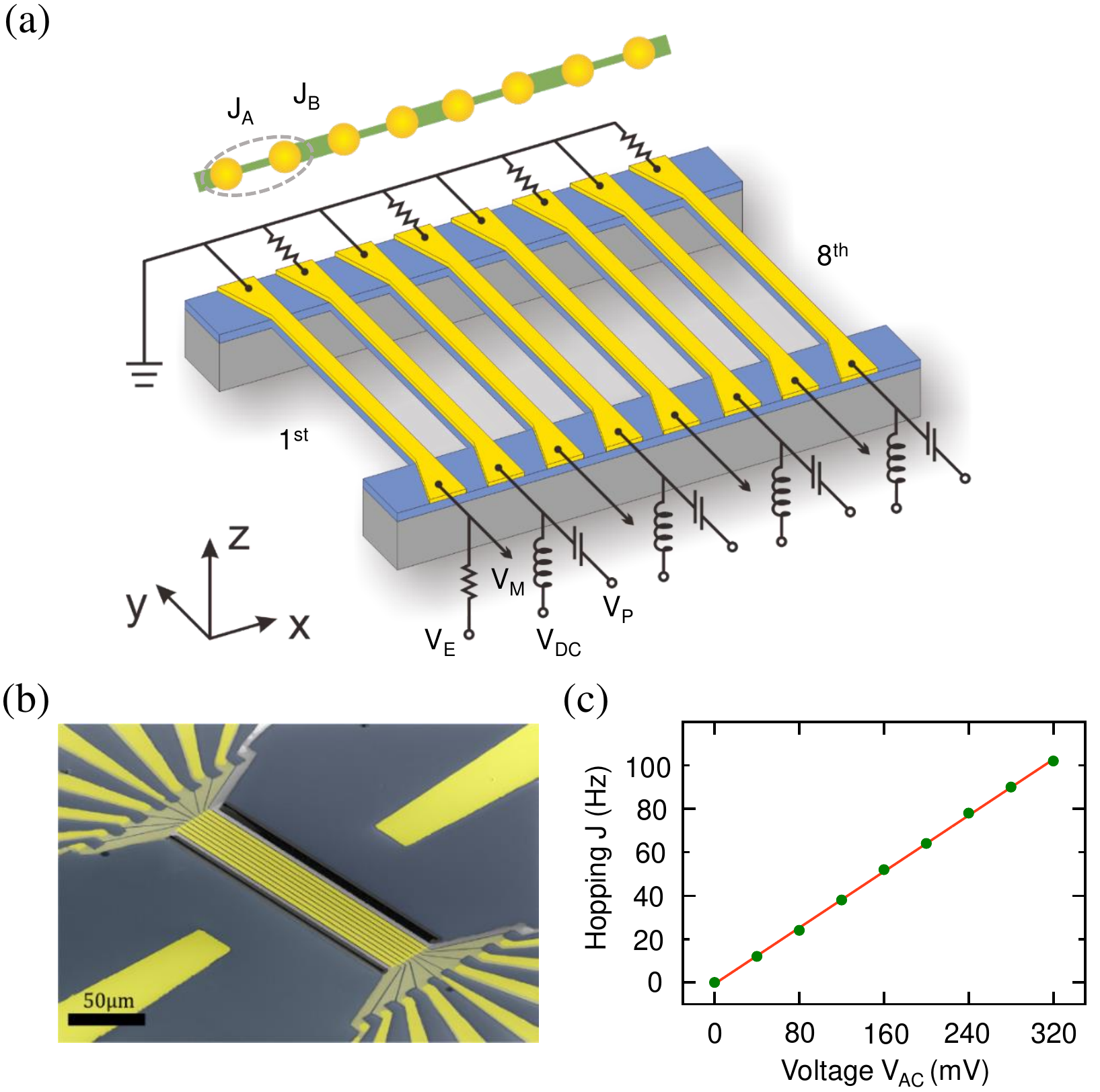}
	\caption{Experimental setup.
		(a) The sample of eight almost identical nanomechanical oscillators (doubly-clamped beams) is fabricated by 100 nm thick silicon nitride (blue) coated with a thin layer of gold. The fundamental out-of-plane modes of these oscillators are used in this work. The nearest-neighbor parametric couplings are realized by applying sum of $V_{\rm DC}$ and $V_{\rm AC}\cos(\omega_{\rm p} t)$ between every two adjacent oscillators. The unit cell of SSH chain is composed of two adjacent oscillators. And the intracell ($J_A$) and intercell ($J_B$) hopping amplitudes can be tuned by parametric couplings. Magnetic field is applied along $x$ axis for excitation ($V_{\rm E}$) and detection ($V_{\rm M}$ with arrows).(b) False-color scanning electron micrography of the sample. (c) The typical hopping amplitude $J$ increases linearly with $V_{\rm AC}$, indicating well controllability of this system.
	}
	\label{setup}
\end{figure}

We study the quench dynamics in a one-dimensional (1D) SSH model~\cite{Su1979}, which describes a 1D super-lattice with two sites per unit cell. The Hamiltonian reads
\begin{equation}
\hat{H}=J_{A}\sum\limits_{j=\mathrm{odd}} |j\rangle\langle j+1|+J_{B}\sum\limits_{j=\mathrm{even}} |j\rangle\langle j+1|+h.c.,
\label{SSHModel}
\end{equation}
where $|j\rangle$ is the state of exciting the $j$-th site, $J_{A}$ and $J_{B}$ are intracell and intercell hopping respectively. Its equilibrium phase diagram and spatial structure are showed in Fig.~\ref{schematic}(a) and the top of Fig.~\ref{setup}(a), respectively.
Its topological aspects can be characterized by a topological invariant - the winding number $\mathcal W$~\cite{Asboth2016}.
In momentum space, the Hamiltonian is given as $\hat{H}(k)=\textbf{d(k)}\bm{\cdot} \bm{\sigma}$ with the Pauli matrices $\bm{\sigma}$.
The trajectory of $\textbf{d(k)}$ does not encircle the origin point when $J_A>J_B$, so $\mathcal W =0$ and the structure is topological trivial.
Conversely,  when $J_A<J_B$, $\mathcal W =1$ and the structure is topological nontrivial, see Fig.~\ref{schematic}(a).
According to the bulk-edge correspondence, there exist topological edge states under the open boundary condition if the bulk invariant is topological nontrivial.
Thus, we can distinguish topological trivial and nontrivial structures by the winding number or the existence of edge states.

We consider the quenches of a zero-energy edge state from the topological Hamiltonian $\hat{H}_i (J_A<J_B)$ to (i) topological trivial Hamiltonian $\hat{H}_{f1} (J_A>J_B)$ and (ii) topological nontrivial Hamiltonian $\hat{H}_{f2} (J_A<J_B)$.
The initial state is prepared at an edge state $|\psi(0)\rangle$ of the topological nontrivial Hamiltonian $\hat{H}_i$.
The evolution of the system follows  $|\psi(t)\rangle=e^{-i\hat{H}_{f}t}|\psi(0)\rangle$ after quenches.
Here, we are interested in the DPT which is related to the Loschmidt amplitude~\cite{Heyl2018},
\begin{equation}
\mathcal {G}(t)=\langle \psi(0)|\psi(t)\rangle = r(t)e^{i\phi(t)},
\label{Loschmidt}
\end{equation}
where $r(t)$ and $\phi(t)=\phi^{\rm dyn}(t)+\phi^{\rm P}$ are the modulus and phase. Herein, $\phi^{\rm dyn}(t)=-\int_0^{t}\langle\psi(s)|\hat{H}_{f}|\psi(s)\rangle ds$ is the dynamical phase and $\phi^{\rm P}$ is the PGP~\cite{Budich2016}.
Both $r(t)$ and $\phi^{\rm P}$ are important in identifying the DPT.
According to the original definition of DPT~\cite{Heyl2013,Heyl2018}, the rate function is derived from Loschmidt amplitude as
\begin{equation}
\lambda(t)=-\frac{1}{N}\ln |\mathcal {G}(t)|^2=-\frac{1}{N}\ln |r(t)|^2,
\end{equation}
which becomes non-analytic at the critical time when DPTs take place.
Meanwhile, the PGP will have a $\pi$ phase jump at the critical time.
In general, the PGP mixes with the dynamical phase, which hinders its direct measurement.
However, we will show that the dynamical phases are absent in our quench scheme.

\subsection{Eliminating the dynamical phase}

To understand the two different quenches, we further decompose the Loschmidt amplitude as the sum of partial Loschmidt amplitudes, i.e.,
\begin{equation}
\mathcal {G}(t)=\sum_{n} \mathcal{G}_{n}(t)=\sum_{n} |\langle \psi_{n}|\psi(0)\rangle|^2 e^{-iE_{n}t},
\label{partial Loschmidt amplitude}
\end{equation}
where $\mathcal{G}_n(t)$ is the partial Loschmidt amplitude with eigenstates $|\psi_{n}\rangle$ and eigenvalues $E_{n}$ of the finial Hamiltonian $\hat{H}_f$.

Due to the chiral symmetry~\cite{Asboth2016}, the initial edge state equally populates all symmetrical pairs of final eigenstates so that the occupation $|\langle \psi_{-m}|\psi(0)\rangle|^2=|\langle \psi_{+m}|\psi(0)\rangle|^2$ for the SSH chain, where $m=1,2,..., N$ and $N$ is the number of unit cell.

In the quench-(i), $\hat{H}_{f}$ is trivial and there is no zero-energy state, the Loschmidt amplitude can be rewritten as
\begin{equation}
\mathcal {G}(t)=2\sum_{m=1}^{N}|\langle \psi_{m}|\psi(0)\rangle|^2 \cos(E_{m}t),
\label{Loschmidt-1}
\end{equation}
Here we have merged the terms of $|{\psi_m}\rangle$ with energy $E_m$ and its chiral symmetry partner $\hat \Gamma |{\psi_m}\rangle$ with energy $-E_m$. For the quench-(ii), the finial Hamiltonian $\hat{H}_{f}$ is nontrival and the Loschmidt amplitude can be rewritten as
\begin{eqnarray}
\mathcal {G}(t)&=&|\langle \psi_{1}|\psi(0)\rangle|^2+|\langle \psi_{-1}|\psi(0)\rangle|^2 \nonumber \\
&+&2\sum_{m=2}^{N}|\langle \psi_{m}|\psi(0)\rangle|^2 \cos(E_{m}t),
\label{Loschmidt-2}
\end{eqnarray}
where $|{\psi_1}\rangle$ and $|{\psi_{-1}}\rangle$ denote the zero-energy eigenstates of $\hat{H}_f$. Either the first or the second term is zero, as a topological edge state only supports one sublattice. Thus, Loschmidt amplitude $\mathcal {G}(t)$ is real so the phase $\phi(t)$ can only be $0$ or $\pi$ all the time for these quenches.

Next, we prove that the dynamical phase $\phi^{\rm dyn}(t)$ is naturally eliminated in all quenches from a topological edge state. Due to the chiral symmetry, the dynamical phase in these processes,
\begin{equation}
\phi^{\rm dyn}(t)=-\sum_{n}|\langle \psi_{n}|\psi(0)\rangle|^2 E_{n}t,
\end{equation}
can be further written as
\begin{equation}
\phi^{\rm dyn}(t)=\sum_{m=1}^{N}\left(|\langle \psi_{-m}|\psi(0)\rangle|^2-|\langle \psi_{m}|\psi(0)\rangle|^2\right) E_{m}t.
\end{equation}
The term in the summation is alway $0$ whether $E_{m}$ is zero or not. So that the dynamcial phase is always zero in the quenches from a topological edge state. The phase $\phi(t)$ which we measure has only the geometric part $\phi^{\rm P}(t)$. Therefore, when the Loschmidt amplitude $\mathcal{G}(t)$ goes through zeros, the DPT takes place while the PGP from 0 jump to $\pi$, see Fig.~\ref{schematic}(b).

The quenches described above are clearly showed on the complex planes in Fig.~\ref{schematic}(c) and Fig.~\ref{schematic}(d). Wherein, each pair of partial Loschmidt amplitudes $\mathcal{G}_n(t)$ in Eq.\eqref{partial Loschmidt amplitude} with $E_{+m}$ and $E_{-m}$ rotate around the origin with same length ($|\langle \psi_{\pm m}|\psi(0)\rangle|^2$) and frequency ($|E_{\pm m}|$) but opposite direction. And there is always a symmetrical part for $\mathcal {G}_{n}$ at a given $t$, as $E_{-m}=-E_{m}$ in the energy bands of SSH model. The summation of all these vectors gives a total vector representing the Loschmidt amplitude (dashed arrows) in real axis. Thereby the PGP is either $0$ or $\pi$. In the sudden quench from the topological nontrivial to trivial phases, all pairs of vectors rotate around the origin with non-zero frequencies and the total vector can change its direction at the critical times, see Fig.~\ref{schematic}(c). At the moment $t_c$, the non-analytic behavior of the rate function appear and PGP jumps $\pi$ simultaneously [Fig.~\ref{schematic}(b)].
In the sudden quench from one topological nontrivial to another nontrivial phases, if the Loschmidt amplitude is dominated by the non-rotating vector pair with zero energy ($|E_{\pm 1}|=0$), the total vector cannot goes through zero to change direction, see Fig.~\ref{schematic}(d).
There is no DPT in this situation and PGP remains zero.

However, there is no guarantee that the non-rotating vector always dominates, especially when the nontrivial phases are quite close to the phase boundary.
Indeed, we find accidental DPTs in the quench from one topological nontrivial to another nontrivial phases near the equilibrium phase boundary.
The relation between DPTs and the underlying equilibrium phases will be analysed later.

\section{Realization of nanomechanical SSH lattices}
We realize the 1D tight-binding SSH model by a reconfigurable mechanical oscillator lattice in experiment, see Fig.~\ref{setup}(a) and Fig.~\ref{setup}(b). The lattice consists of eight almost identical oscillators (doubly clamped beams). Every unit cell of SSH model is composed of two adjacent oscillators.
The intracell and intercell hopping amplitudes are controlled by different nearest-neighboring parametric couplings, which are realized by applying voltages $V_{\rm DC}+V_{\rm AC}\cos(\omega_{\rm p}t)$ between adjacent oscillators~\cite{Huang2013}.
The parametric coupling increases linearly with $V_{\rm AC}$, which indicates this system can be reconfigured by tuning voltages, see Fig.~\ref{setup}(c).

\begin{figure*}\centering
	\includegraphics[width=2\columnwidth]{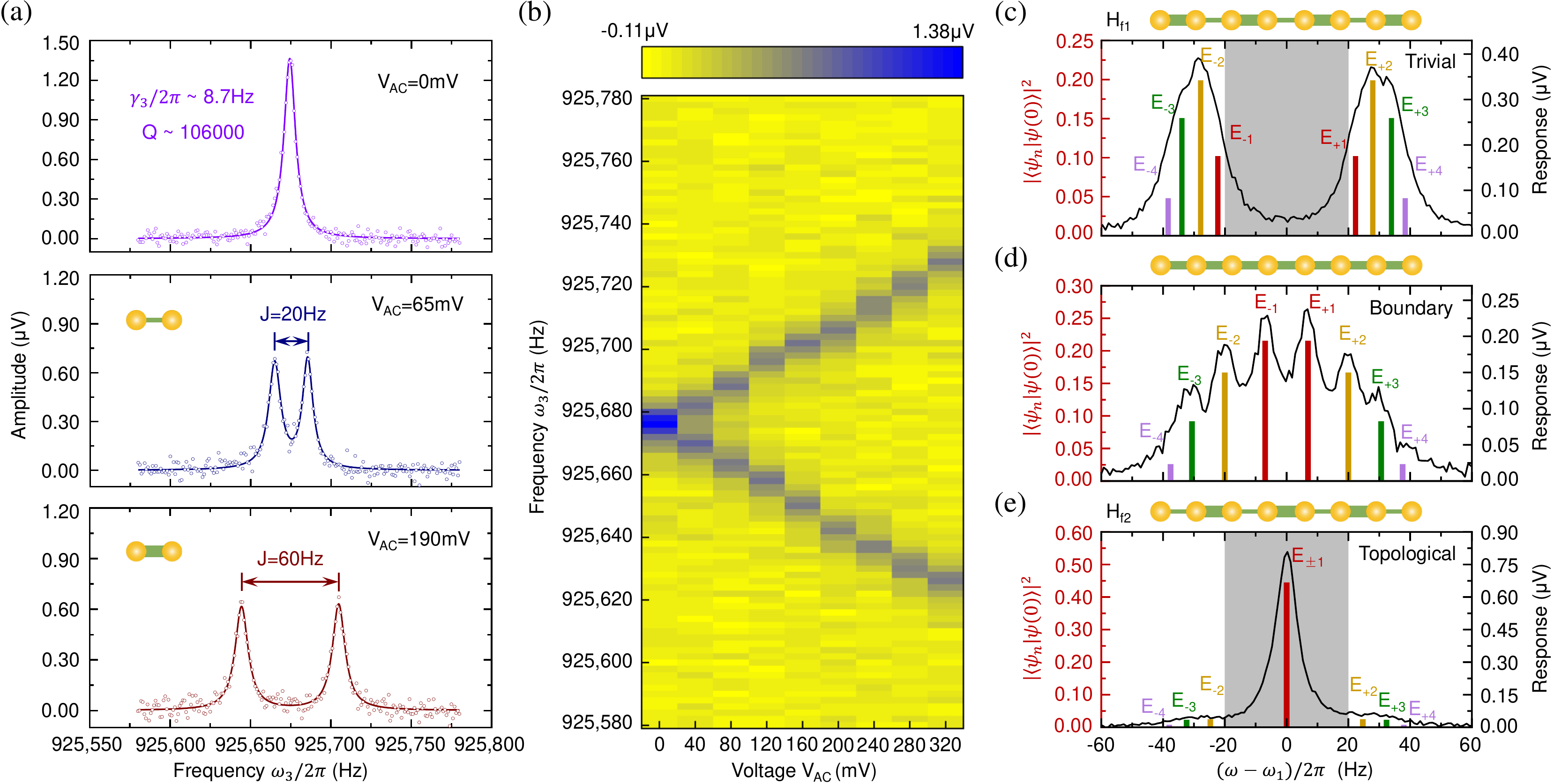}
	\caption{Configuring different topological structures in situ.
		(a),(b) Typical parametric coupling between two adjacent oscillators. The signal is measurement at the 3rd oscillator. And the pump $V_{\rm DC}=4{\rm V}$ combined with $V_{\rm AC}\cos(\omega_{\rm p} t)$ is applied in the 2nd oscillator. Consequently, the arisen splits denote tunable coupling strengths, which are correspond to different hopping amplitudes in SSH model. (a) The quality factor of mechanical mode can reach about $1 \times10^5$ at 77 K in vacuum. The strong couplings of 20 Hz and 60 Hz are used to configure two different topological structures. (b) The peak split spectrum when applying different $V_{\rm AC}$. (c)-(e) Response spectrum (black lines) of three different structures of SSH model, which are measured at the edge oscillator. And the theoretical occupations of the initial edge state in final eigenstates are also showed (colored bars). (c) Topological trivial phase $\hat H_{f1}$. The intracell and intercell coupling strengths are 60 Hz and 20 Hz alternately. Two symmetric bands separated by a gap (grey area) are observed. (d) Boundary trivial phase. The coupling strengths are 40 Hz in entire chain. No gap can be observed. (e) Topological phase $\hat H_{f2}$. The intracell and intercell coupling strengths are 20 Hz and 60 Hz alternately. The zero mode in the energy gap is observed.
	}
	\label{Response}
\end{figure*}

The parametric coupling strength is determined as follows.
By applying the voltage $V_{\rm DC}$ and $V_{\rm AC}\cos(\omega_{\rm p}t)$ in the 2nd oscillator to couple the 2nd and 3rd oscillators.
All $V_{\rm DC}$ are chosen as 4 V in this work to keep the frequency difference $\omega_{\rm p}^{j}=\omega_{j}-\omega_{j+1}$ stable.
We measure the frequency response at the 3rd oscillator, where the split between two frequency peaks gives the parametric coupling strength.
Fig.~\ref{Response}(a) shows the frequency responses at three typical AC voltages $V_{\rm AC}$ = 0, 65, 190 mV.
The parametric coupling strength linearly increases with the AC voltage, see the peak split spectrum at different $V_{\rm AC}$ in Fig.~\ref{Response}(b).

The classical motion equations of the coupled oscillators can be fully mapped onto the Hamiltonian of SSH model, see Appendix \ref{SSHMechanical} for more details.
%
By tuning different couplings between adjacent oscillators, we can realize various structures of SSH model. Specifically, we choose an edge state $|\psi(0)\rangle=(1,0,...,0)^{T}$ of a large dimeric topological Hamiltonian $\hat{H}_i$ with $J_A/J_B\rightarrow 0$ as the initial state. Therefore the initial state is prepared at the excitation of edge oscillator.
In Fig.~\ref{Response}, we configure three structures and measure its response spectrum at the edge oscillator.
Fig.~\ref{Response}(c) shows a topological trivial chain and it is used to simulate the Hamiltonian $\hat H_{f1}$ in Fig.~\ref{schematic}(a), with the coupling strengths being 60 Hz and 20 Hz alternately. There is a band gap (gray area) between two frequency bands in the spectrum.
Fig.~\ref{Response}(d) shows a trivial phase at the boundary between topological and trivial phase.
There is a continuous band with all the coupling strength being 40 Hz.
In Fig.~\ref{Response}(e), we realize a topological phase by choosing 20 Hz and 60 Hz alternately. It is used to simulate the Hamiltonian $\hat H_{f2}$ in Fig.~\ref{schematic}(a).
There is a distinct zero mode in the band gap compared with Fig.~\ref{Response}(c). All the spectrum profiles match the occupation of eigenstates at the edge oscillator, $|\langle \psi_n|\psi (0)\rangle|^2$.
Although eight peaks are not distinguishable in these spectrums due to the dissipation of oscillators, the appearance of zero mode is remarkable distinction between the topological and the trivial.

\section{Observation of dynamical phase transitions}

In this section, we show how to simulate and observe DPTs in parametrically coupled nanomechanical oscillators. Because of the mode overlaps in different SSH structures, we can obseve DPTs via the dynamical evolution of initial edge excitation. The rate function and the PGP can be measured via the normalized amplitude and vibration phase of edge oscillator. The experimental results indicate that DPTs are robust against the disorder and the fluctuations of critical time increase with the disorder strength.


\subsection{Measurement of rate function and PGP}
\begin{figure*}\centering
	\includegraphics[width=2\columnwidth]{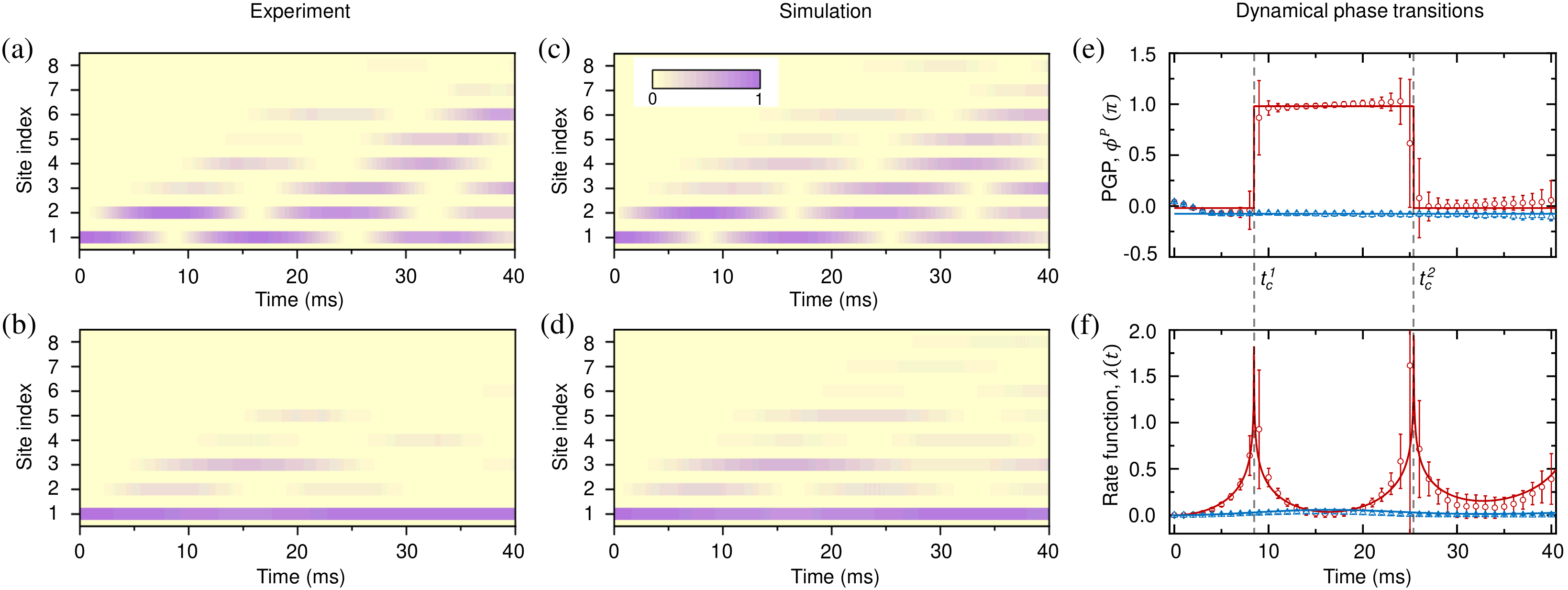}
	\caption{Observation of dynamical phase transitions. (a) Experimental result of the dynamical evolution from an edge excitation to the trivial SSH structure. Initial edge excitation propagates over the entire chain. (b) Experimental result of the dynamical evolution from an edge excitation to the topological nontrivial SSH structure. The edge oscillator dominates the time-evolution. (c) and (d) Numerical simulation counterpart of (a) and (b), respectively. All amplitudes showed in (a)-(d) are normalized at every moment. (e) The PGPs in quench-(i) and quench-(ii), which are observed from the demodulating phase of the edge oscillator. The PGP jumps $\pi$ at each critical time in quench-(i), but remains unchanged in quench-(ii). The initial values of PGP are non-zeros constants because of measuring circuit. (f) The rate functions for quench-(i) and quench-(ii), which are derived from the normalized amplitude of the edge oscillator in (a) and (b), the non-analytic behavior are obvious at DPTs. All solid lines in (e) and (f) are theoretical results. All error bars denote statistical confidence of one standard deviations.}
	\label{RatePGP}
\end{figure*}

In experiment, quenches of topological edge state in different topological phases are realized by time-dependent tuning different couplings between adjacent oscillators.
Specifically, we use different pulse sequences to configure structures in time domain, see Appendix~\ref{MeasurementScheme} for more details.
The preparation initial edge state is realized by applying a sinusoidal wave to the first oscillator.
At time $t=0$, we turn on all the coupling voltages between every two adjacent oscillators to realize SSH structures.
As the time evolves, we measure the vibration of each oscillator with a lock-in amplifier detection.
Here, we assume that all oscillators have the same decay rates in the short time of evolution, as the strong coupling strengths are much larger than the decay rates and the differences between the oscillators are small.
Therefore, all the amplitudes can be normalized at every moment.
We show the normalized vibration of each oscillator in 40 ms evolution, see Fig.~\ref{RatePGP}(a) for quench-(i) and Fig.~\ref{RatePGP}(b) for quench-(ii).
%
In quench-(i), the initial edge excitation propagates over the entire chain under the trivial structure.
%
In quench-(ii), the initial edge excitation dominates the time-evolution and most of the amplitudes localize at the edge site under the topological structure.
We also numerically simulate the two different quenched dynamics without any fitting parameter, see Fig.~\ref{RatePGP}(c) and Fig.~\ref{RatePGP}(d), respectively.
The experimental results and numerical simulations are in good agreement.

Apart from the amplitude of every oscillator, one can directly extract the phase from the lock-in phase-sensitive detection. The vibration phase of the edge oscillator is analogous to the PGP of the Loschmidt amplitude.
%
In quench-(i), we observe the PGP jumps twice within 40 ms evolution, identifying the DPTs exactly as the theoretical prediction, see the red circles in Fig.~\ref{RatePGP}(e).
The large variances of PGP around the critical times $t_{c}^{1}$ = 8.45 ms and $t_{c}^{2}$ = 25.40 ms [see Appendix \ref{CriticalTime}] also indicate the zero of Loschmidt amplitude.
In the quench-(ii), we observe the PGP keeps unchanged in 40 ms evolution where no DPT happens,  see the blue triangles in Fig.~\ref{RatePGP}(e).
The initial non-zero PGP results from the external electric circuits, and it does not affect the jump behavior.

The rate function can also be used to diagnose the DPTs.
We obtain the rate function from the normalized amplitude of the first oscillator in the two quenches, and have observed non-analytic behaviors of the rate function when the system was quenched across an underlying topological phase transition, see the red circles with error bar in the Fig.~\ref{RatePGP}(f).
The non-analytic behaviors directly verify the DPTs.
For the quench-(ii) in the same topological phase, the rate function is always analytic in 40 ms evolution, which suggests that there is no appearance of DPTs, see the blue triangle with error bar in the Fig.~\ref{RatePGP}(f).
All the experimental results are measured by averaging 500 times, and they are in great agreement with the theoretical results [solid lines in Fig.~\ref{RatePGP}(e) and Fig.~\ref{RatePGP}(f)] obtained by solving the Schr\"{o}dinger evolution equation.

\subsection{Robustness of dynamical phase transitions}
To investigate the robutness of DPTs, we introduce disorder into hopping strength as $J_{A(B)}+\delta J_j$, where $\delta J_j$ is a random number in $[-\Delta, \Delta]$.
It is easily to check that the systems still preserve the chiral symmetry.
Since the jump of PGP directly identifies DPT, we explore the responses of PGP as the disorder strength increases.

\begin{figure}
	\includegraphics[width=1\columnwidth]{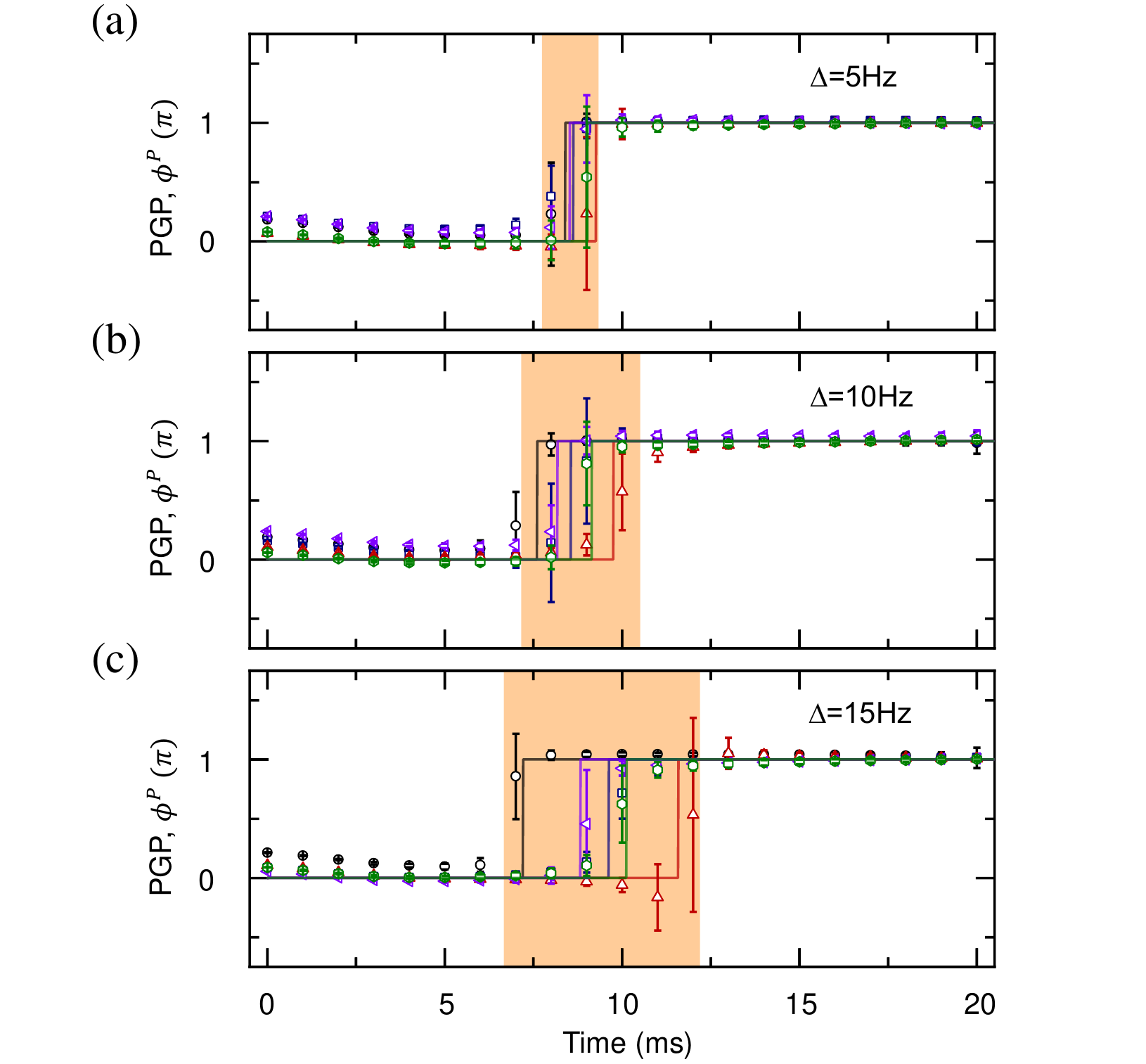}
	\caption{Robustness of DPTs against structure imperfections. DPTs are diagnosed by PGP in quench-(i) with three different disorder strengths: (a) $\Delta$ = 5 Hz, (b) $\Delta$ = 10 Hz and (c) $\Delta$ = 15 Hz. There are 5 disorder samples in each disorder strength. All error bars denote statistical confidence of one standard deviations. All solid lines in (a)-(c) are numerical results. The fluctuations of the first critical time (shadow region) are obtained by numerical simulation.
	}
	\label{robutness of DPTs}
\end{figure}

In experiment, we follow the similar procedures in quench-(i). The parameters are chosen the same as those in quench-(i) except for additional disorders $\delta J_j$, which are controlled by different voltage $V_{\rm AC}(t)$ between adjacent oscillators.
We randomly generate five disorder samples for each disorder strength, and observe the PGP as a function of time, see Figs.~\ref{robutness of DPTs}(a),~\ref{robutness of DPTs}(b) and ~\ref{robutness of DPTs}(c) with the corresponding disorder strength $\Delta$ = 5 Hz, 10 Hz and 15 Hz.
We find that DPTs happen in all those cases.
As the disorder strength increases, the fluctuations of the first critical time increase, see the shadow region obtained by numerical calculations.
As the time $t$ goes further, it is natural that the fluctuations of subsequent critical times of latter DPTs also increase.

\section{Relation between DPTs and equilibrium phase boundary}
Whether it is possible to define the phase boundary by the appearance of DPTs, similar to the equilibrium phase boundary showed on the dashed line in Fig.~\ref{schematic}(a)? The size effects need to be taken care of before solving this question.
Different from DPTs in the momentum space (periodic boundary condition) ~\cite{Heyl2013,Heyl2018,Budich2016,Vajna2015,ZHuang2016,Schmitt2015,Flaschner2018,Qiu2018,Bhattacharya2017,Heyl2017}, our model is set in real space with open boundary condition. Considering the quenches of limited system size, some trivial zeros of Loschmidt amplitude will present at sufficient evolution time. However, as these trivial zeros will disappear as the system size increases, they cannot be regarded as DPTs.

\begin{figure}
	\includegraphics[width=1\columnwidth]{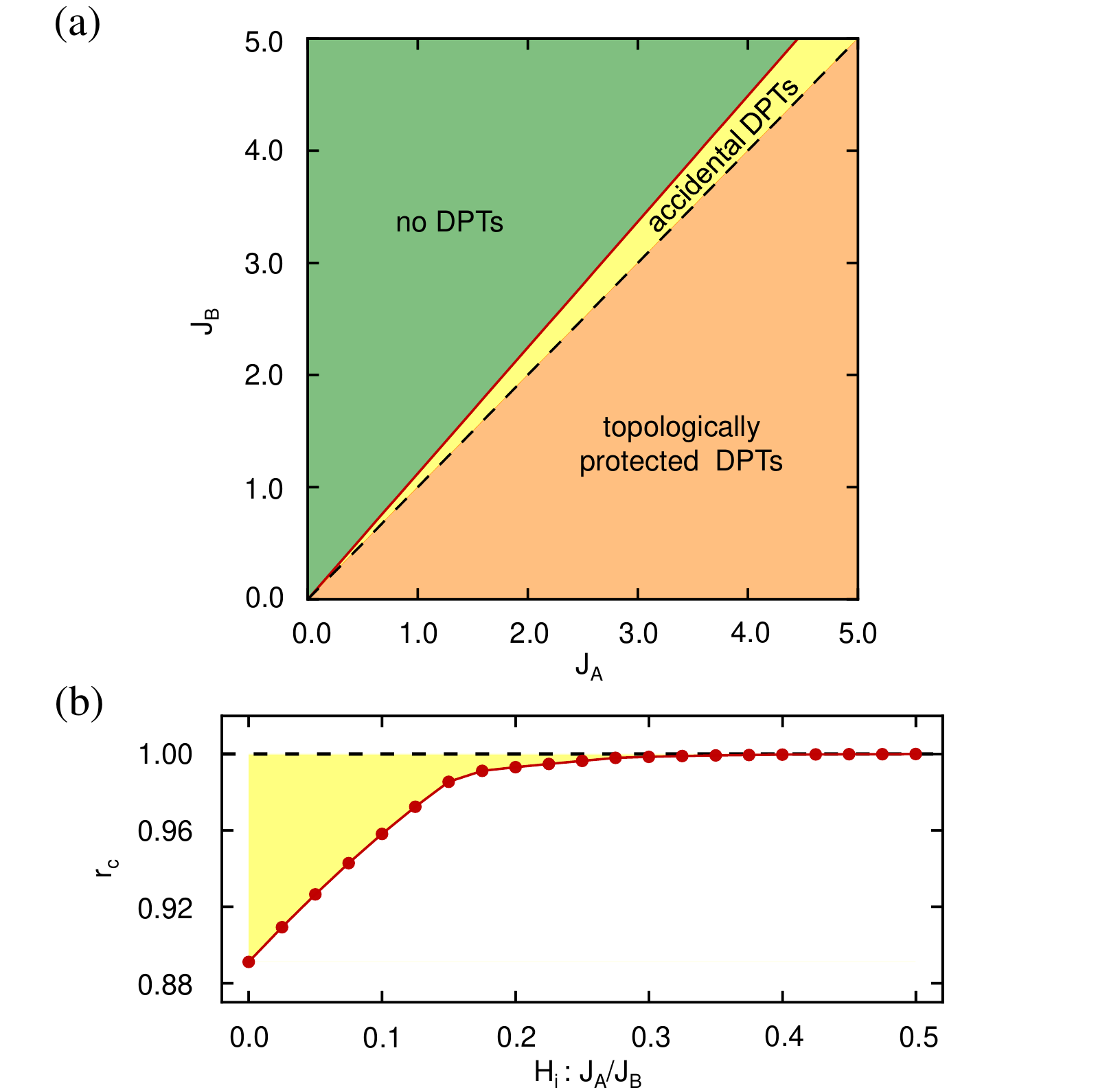}
	\caption{Phase diagram of dynamical phase transitions. (a) Simulation result for the initial edge state of $\hat{H}_i: J_A/J_B\rightarrow 0$. The red solid line $J_A/J_B=r_c\approx 0.8911$ represents the dynamical phase boundary of DPTs and the black dashed line represents the underlying topological phase boundary. Accidental DPTs occur at the yellow region between dynamical and equilibrium phase boundary. (b) The dynamical phase boundary $r_c$ for different initial nontrivial Hamiltonian $\hat{H}_i$.
	}
	\label{dynamical phase boundary}
\end{figure}

To understand the relation between DPT and underlying equilibrium phase boundary, we numerically calculate the phase diagram of the DPTs in the whole parameter space. See Fig.~\ref{dynamical phase boundary}(a), we choose 80 sites and finite time to study the quenches from the edge state of Haimltonian $ \hat{H}_i$ with $J_A/J_B\rightarrow 0$.
The systems evolve under the final Hamiltonian with intercell coupling $J_A$ and intracell coupling $J_B$.
The phase diagram of DPTs is diagnosed by the jump of PGP.
The dynamical phase boundary of DPTs is approximately given as $J_A/J_B=r_c\approx 0.8911$, see the red solid line.
The solid line suggests DPTs will happen when $J_A>r_c J_B$ but will not when $J_A<r_c J_B$.
The phase boundary of underlying topological phase is given as $J_A=J_B$, see the black dashed line.
Apparently, the phase boundary of DPTs departs from the phase boundary of underlying topological phases.

If the quench takes place in the same topological phases, the DPTs may happen near the equilibrium phase boundary, see the yellow region in Fig.~\ref{dynamical phase boundary}(a). As reported in previous studies, these DPTs are named as accidental (or topologically non-protected) DPTs and they require fine-tuning of the Hamiltonian~\cite{Heyl2018,Vajna2015,ZHuang2016}.
Indeed, the accidental DPTs are unstable as the initial state changes. We numerically calculate the quenches from different initial nontrivial Hamiltonians $\hat{H}_i (J_A/J_B<1)$. As showed in Fig.~\ref{dynamical phase boundary}(b), dynamical phase boundary $r_c$ will approach the equilibrium phase boundary $ J_A/J_B =1$ and the accidental DPTs will disappear with the increase of $J_A/J_B$.

In contrast to accidental DPTs, topologically protected DPTs  always occur when the system is quenched across the underlying equilibrium phase boundary, which is consistent with previous studies in momentum space~\cite{Budich2016,Vajna2015,ZHuang2016,Sedlmayr2018}. Thereby topologically protected DPTs rather than accidental DPTs closely connect with the equilibrium phase boundary in this work.

\section{Summary and discussions}
In summary, by using a reconfigurable nanomechanical lattice, we simulate DPTs in SSH model and directly observe PGP and rate function after a sudden quench from a topological edge state. We also prove the robustness of DPTs in presence of hopping disorders, and give a connection between topologically protected DPTs and equilibrium phase diagram. The theoretical scheme present here that eliminating the dynamical phase can be generalized to some other models which process chiral symmetry or particle-hole symmetry, such as the off-diagonal AAH model~\cite{Ganeshan2013} and the Kitaev model~\cite{kitaev2001}. Our experiment shows that DPTs can be observed by directly measuring quantitative PGP. The results suggest the PGP can be seen as a dynamical order parameter in analogy to the order parameter of phase transitions, and can be used as an accurate method to diagnose DPTs even for the low sample rate in experiment.

For a topological system, there exists bulk-edge correspondence in the equilibrium where the number of edge states corresponds to the bulk topological invariants~\cite{Hatsugai1993}. There some interests in the bulk-boundary correspondence for dynamical phase transitions~\cite{Sedlmayr2018}. Our work shows that quenching across topological nontrivial phases is a sufficient and necessary condition for the occurrence of topologically protected DPTs in SSH model with open boundary conditions. Exploring the bulk-edge relations in quenched dynamics will be an interesting and important problem deserving further study.

In addition, the reported nanomechanical systems simultaneously possess three features: multiple modes, high quality factor, and dynamically tunable couplings, which were only partially realized previously~\cite{Weig2013,Weig2017,Okamoto2013,Huang2013,Huang2016}. This is not only a technical advance but also a versatile platform for non-equilibrium physics and topology. It can be utilized to study more complicated phenomena including interplay between topology, dissipation and nonequilibrium dynamics~\cite{Zhou2017}.

\textit{Note added:} During the preparation of this manuscript, we find three other works~\cite{GuoXY2018,Wangk2018,Smale2018} also report the observations of DPTs in different systems. In the work of photonic quantum walks~\cite{Wangk2018}, the PGP is indirectly obtained by deducting the dynamical phase from the total phase of the Loschmidt amplitude.

\begin{acknowledgements}
We thank Fazhan Shi, Changkui Duan and Ya Wang for discussion on the manuscript. This work was supported by the National Key R\&D Program of China (Grant No. 2018YFA0306600), the CAS (Grants No. GJJSTD20170001 and No. QYZDY-SSW-SLH004), and Anhui Initiative in Quantum Information Technologies (Grant No. AHY050000), and the National Natural Science Foundation of China (NNSFC) under Grants No. 11874434, 11574405, 11704420 and 11675163.
\end{acknowledgements}

\appendix
\makeatletter

\section{Theory}
\subsection{SSH model represented by mechanical oscillators} \label{SSHMechanical}
The motion of $2N$ classical oscillators with nearest-neighbor parametric coupling can be described as:
\begin{equation}
m_j \ddot{z}_j+k_jz_j=L_{j}(t)(z_{j+1}-z_{j})+L_{j-1}(t)(z_{j-1}-z_{j}).
\label{classical}
\end{equation}
Here $m_j$ and $k_j = m_j \omega_j^2$ are the corresponding effective mass and spring constant,
\begin{equation}
L_{j}(t) = \eta_{j} {\cos}(\omega_{\rm p}^{j} t)
\label{coupling}
\end{equation}
is the time-dependent coupling and $\eta_{j}$ is the coupling strength. For the problem concerned in this work, $\omega_{\rm p}^j$ fulfils the frequency conversion condition:
\begin{align}
\label{}
\omega_{\rm p}^j= (\omega_{j}-\omega_{j+1}) \ll \omega_{j}.
\end{align}
Using a slowly varying complex amplitudes $\psi_j$ in $z_j=A_{j}\Re(\psi_je^{i\omega_j t})$ and ignoring the terms $\ddot{\psi}_j$, we get
\begin{eqnarray}
&& i \dot{\psi}_j+\frac{L_{j}(t)+L_{j-1}(t)}{2m_{j}\omega_{j}}\psi_j \nonumber \\
&&= \frac{L_{j-1}(t)}{2m_j\omega_j}\frac{A_{j-1}}{A_{j}}e^{i\omega_{\rm p}^{j-1}t}\psi_{j-1}+
\frac{L_j(t)}{2m_j\omega_j}\frac{A_{j+1}}{A_{j}}e^{-i\omega_{\rm p}^jt}\psi_{j+1}, \nonumber \\
\label{classical-2}
\end{eqnarray}
where $A_{j}=\sqrt{{\hbar \omega_j}/{(2k_j)}}$ is constant.

According to \eqref{coupling}, \eqref{classical-2} can be simplified
by adopting the rotating wave approximation as,
\begin{equation}
i \dot{\psi}_j=
\frac{A_{j-1}}{A_{j}}\frac{\eta_{j-1}}{4m_j\omega_j}\psi_{j-1}+
\frac{A_{j+1}}{A_{j}}\frac{\eta_j}{4m_j\omega_j}\psi_{j+1}.
\label{classical-3}
\end{equation}

We introduce $|\psi\rangle=\frac1L(\psi _1,\psi _2,\psi _3,...,\psi_{2N})^T$ as a single-particle wavefunction in a 1D potential. The $L$ is a normalizing constant, $L=\sum_j|\psi_j|^2$. Thus, the equation of a mechanical wave \eqref{classical-3} is fully mapped onto the Schr\"{o}dinger equation (or see~\cite{Briggs2012,Briggs2013}), $i\frac{d }{dt}|\psi\rangle =\hat{H}|\psi\rangle$.
%
By using $|j\rangle$ to denote $\psi_j/L$, the Hamiltonian $\hat{H}$ is rewritten as,
\begin{equation}
\hat{H}= \sum_{j}^{2N}\frac{\eta_{j}}{4}\sqrt{\frac{\omega_{j} \omega_{j+1}}{k_{j}k_{j+1}}}(|j\rangle \langle j+1|+|j+1\rangle \langle j|).
\label{Hamiltonian-H}
\end{equation}

So if we set $\eta_{j}\sqrt{\omega_{j}\omega_{j+1}/{k_{j}k_{j+1}}}/4=J_A (J_{B})$ for $j$ odd(even), $\hat{H}$ represents a SSH model~\cite{Su1979} with $N$ unit cells, i.e.,
\begin{equation}
\hat{H}_{\rm SSH}=J_{A}\sum\limits_{j=2m-1} |j\rangle\langle j+1|+J_{B}\sum\limits_{j=2m} \left|j\rangle\langle j+1|+h.c.\right.
\end{equation}
In this way, we use the coupled nanomechancial oscillators to configure the SSH Hamiltonian.

We did not consider the dissipation in above process. Actually, this is reasonable when the decay rate $\gamma_j$ of each beam is far less than the coupling strength. Furthermore, the decay effect can also be ignored by renormalizing the amplitudes at each moment when all beams have almost the same decay rate $\gamma_j=\gamma $ $(j=1,2,...,8)$. Based on these considerations, the decay effect is neglected in this work.

\subsection{The chiral symmetry in the SSH model} \label{ChiralSSH}
We introduce two orthogonal operators $\hat{P}_O$ and $\hat{P}_E$, which respectively project to the odd sublattice and even sublattice. $\hat{P}_O$ and $\hat{P}_E$ satisfy the following relations,
\begin{eqnarray}
&&\hat{P}_O=\sum_{j=1}^N|2j-1\rangle\langle 2j-1|, \nonumber \\
&&\hat{P}_E=\sum_{j=1}^{N}|2j\rangle\langle 2j|,\nonumber \\
&&\hat{P}_O+\hat{P}_E=\hat I, ~\hat{P}_O\hat{P}_E=0.
\end{eqnarray}
The chiral symmetry operator in SSH model is represented by
\begin{align}
\hat\Gamma=\hat{P}_O-\hat{P}_E=\hat\Gamma^\dag,
\end{align}
and it anti-commutes with the Hamiltonian, i.e., $\hat\Gamma^\dag\hat{H}\hat\Gamma=-\hat{H}$~\cite{Asboth2016}. Thus, the SSH model has chiral symmetry, which leads to a symmetric energy spectrum. In other words, there always exists a chiral symmetric eigenvalue $E_{-m}$ for each eigenvalue $E_{+m}$, and $E_{-m}=-E_{+m}$ ($m=1,2,...,N$).
Besides, the eigenstates are $|{\psi_{+m}}\rangle$ with energy $E_{+m}$ and $ |{\psi_{-m}}\rangle=\hat \Gamma |{\psi_{+m}}\rangle$ with energy $E_{-m}$. This is easy to know that
\begin{eqnarray}
&&\hat{H}|{\psi_{-m}}\rangle=\hat H \hat \Gamma|{\psi_{+m}}\rangle \nonumber \\
&=&-\hat \Gamma \hat{H}|{\psi_{+m}}\rangle=-E_{+m}\hat \Gamma |{\psi_{+m}}\rangle \nonumber \\
&=&-E_{+m}|{\psi_{-m}}\rangle=E_{-m}|{\psi_{-m}}\rangle.
\end{eqnarray}

For $E_n\neq 0$, the eigenstates $|{\psi_n}\rangle$ and $\hat \Gamma|{\psi_n}\rangle$ are orthogonal, which implies that each non-zero energy eigenstate equally supports on both sublattices~\cite{Asboth2016},
\begin{align}
\langle{\psi_n}|\hat{P}_O|{\psi_n}\rangle-\langle{\psi_n}|\hat{P}_E|{\psi_n}\rangle=\langle{\psi_n}|\hat \Gamma|{\psi_n}\rangle=0.
\end{align}

For $E_{n}=0$, the zero-energy eigenstate supports only one sublattice. As the following shows, $\langle{\psi_n}|\hat{P}_O-\hat{P}_E|{\psi_n}\rangle=\pm1$, so either of the projections into the two sublattices vanishes alternatively.
\begin{align}
\hat{H}|{\psi_n}\rangle=0,~\hat{H}\hat \Gamma|{\psi_n}\rangle=0, ~\hat \Gamma|{\psi_n}\rangle=\pm|{\psi_n}\rangle.
\end{align}

\subsection{Critical time} \label{CriticalTime}
We choose a topological edge state of a large dimeric ($J_{A}\ll J_{B}$) SSH model as the initial state in the quenches. The great advantage is that we can directly measure the PGP from the edge site. The trivial phase $\hat{H}_{f1}$ ($J_{A}>J_{B}$) and topological phase $\hat{H}_{f2}$ ($J_{A}<J_{B}$) are realized by coupling strength of 20 Hz and 60 Hz alternately in our system.

According to Eq. \eqref{Loschmidt-1}, we can easily get the theoretical critical times in the quench (from $\hat{H}_{i}$ to $\hat{H}_{f1}$). As showed in Fig.~\ref{RatePGP}(e) and Fig.~\ref{RatePGP}(f), the two critical times are $t_{c}^{1}$ = 8.45 ms and $t_{c}^{2}$ = 25.40 ms, respectively.

\section{Experiment}

\subsection{Fabrication of the sample} \label{Fabrication}
The sample is fabricated as doubly clamped beams with 200 $\mu$m long on wafers composed of a silicon substrate with 100 nm layer of high-stress (1 GPa) silicon nitride using the method of low pressure chemical vapor deposition. Electrodes and the wide beam (3 $\mu$m) are defined by standard ultra-violet lithography. All beams and electrodes are thickened by 10 nm Au. The distance between each two adjacent beams is narrowed to 500 nm by means of e-beam lithography and ion beam etching (remove metal layer). Reactive ion etching is then employed to eliminate silicon nitride layer. Finally, the mechanical beams are suspended by KOH wet etch. These beams are numbered from 1 to 8. The parameters of the first out-of-plane vibrational mode of every beam in experiments are listed in Table~\ref{tab:table1}.
\begin{table}[h]
	\caption{\label{tab:table1} The first out-of-plane vibrational mode of oscillators}
	\begin{ruledtabular}
		\begin{tabular}{cccc}
			NO.  & Frequency/kHz  &Quality factor     \\ \hline
			1    & 907.184    &106300 \\
			2       & 905.980          &91700 \\
			3       & 923.843         &101000 \\
			4       & 893.665         &77400 \\
			5       & 922.695        &105100 \\
			6       & 905.627         &71400 \\
			7       & 918.246        &119600 \\
			8       & 873.976       &84000 \\
		\end{tabular}
	\end{ruledtabular}
\end{table}

\subsection{Parametric Couplings} \label{ParametricCouple}
To generate the electrostatic forces for parametric couplings between adjacent beams, a DC voltage $V_{\rm DC}$ and an AC voltage $ V_{\rm AC}^{ j}(t) $ are applied between $ j$-th and $(j+1)$-th beam, the coupling as the following form can be generated~\cite{Huang2013}.
\begin{equation}
L_{j}(t)= \frac{\partial^2 C_{j}(\delta z_{j})}{ \partial ^{2}\delta z_{j} }V_{\rm AC}^{j}(t)V_{\rm DC}
\end{equation}
Here, $C_{j}(\delta z_{j})$ is the effective capacity between the oscillators and $\delta z_{j} = z_{j}-z_{j+1}$. The form of AC voltage is $V_{\rm AC}^{j}\cos(\omega_{\rm p}^{j}t)$ with $\omega_{\rm p}^{j}=\omega_{j}-\omega_{j+1}$ and $V_{\rm AC}^{j}\ll V_{\rm DC}$ in experiment.

\begin{figure}[h]
	\includegraphics[width=0.8\columnwidth]{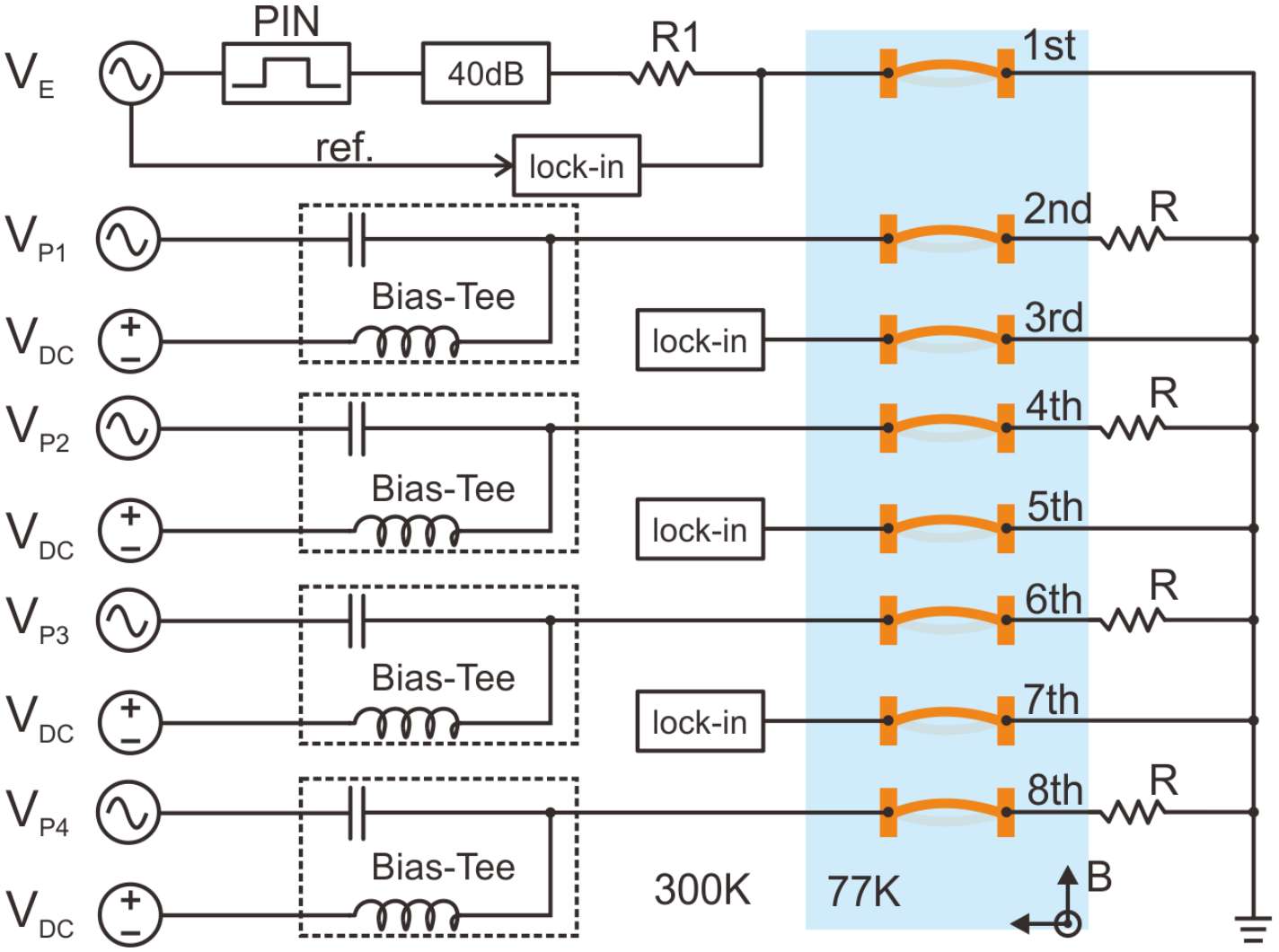}
	\caption{Measurement circuit of odd oscillators.}
	\label{odd oscillators}
\end{figure}

\begin{figure}[h]
	\includegraphics[width=0.8\columnwidth]{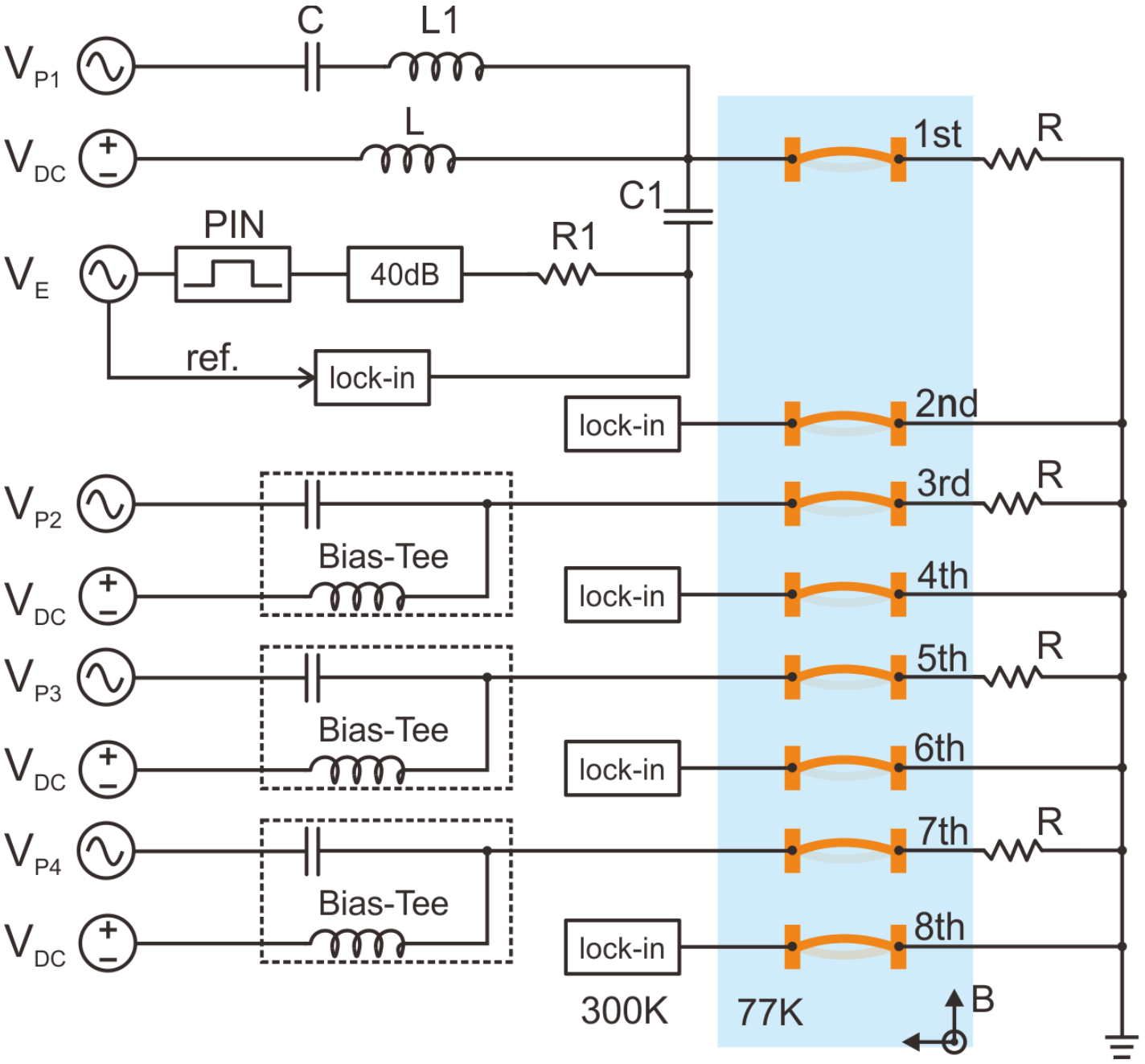}
	\caption{Measurement circuit of even oscillators.}
	\label{even oscillators}
\end{figure}

The DC voltage and AC voltage are combined by a bias-tee, as showed in Fig.~\ref{odd oscillators} and Fig.~\ref{even oscillators} where $V_{P}$ is a sum of the neighborhood AC coupling voltage. The typical value of AC voltage used are listed in Table~\ref{tab:table2} and Table~\ref{tab:table3}.

The effective coupling strength is confirmed by measuring frequency spectrum of the response. As showed in Fig.~\ref{Response}(a) and Fig.~\ref{Response}(b), we measure the response of one oscillator coupled with another adjacent one. The split of peak directly describes coupling strength. In this work, the data of Fig.~\ref{Response}(a) and Fig.~\ref{Response}(b) were measured at another similar sample because previous one was damaged by accident.

The frequency response of entire coupled chain is measured at edge oscillators, corresponding to topological trivial phase and nontrivial phase of SSH model, see Fig.~\ref{Response}(c) and Fig.~\ref{Response}(e). Although eight peaks are not distinguishable in these spectrums because of oscillators' dissipation, the appearance of zero mode is remarkable distinction between topological and trivial. We also measure the response under the same coupling strengths, see Fig.~\ref{Response}(d).

\subsection{Measurement Scheme} \label{MeasurementScheme}
The sample is placed in a vacuum of 3$\times$$10^{-6}$ Pa and cooled to 77 K for the stable frequency and high quality factor. The other measurement circuit at room temperature and all measurements followed the standard magneto-drive method~\cite{Cleland1999}. The standard lock-in amplifiers (The Zurich Instruments HF2LI) are used to measure the frequency-domain spectrums and time-domain measurements. All the frequency generation setups are referencing an external atomic clock to ensure the frequency stability.

We note that the all the oscillators work in linear regime. This can be controlled by the excited strength at the first oscillator. For any oscillator in linear regime, the relation between amplitude $|z(\omega)|$ of the vibration and the measured voltage amplitude $|V(\omega)|$ is
\begin{align}
\label{}
|z(\omega)|=|V(\omega)|/\xi BL \omega
\end{align}
with $\xi \approx 0.83 $ the shape factor of the first vibrational mode.

For the parametric coupling, The time-domain measurement is divided into two parts, as showed in Fig.~\ref{odd oscillators} and Fig.~\ref{even oscillators}. We measure the amplitudes of odd oscillators when coupling voltages are applied to even oscillators and measure the amplitudes of even oscillators when coupling voltages are applied to odd oscillators. The amplitude of the first oscillator is monitored to ensure the initial excited strength is the same in both parts.

\begin{figure*}[h]
	\includegraphics[width=2\columnwidth]{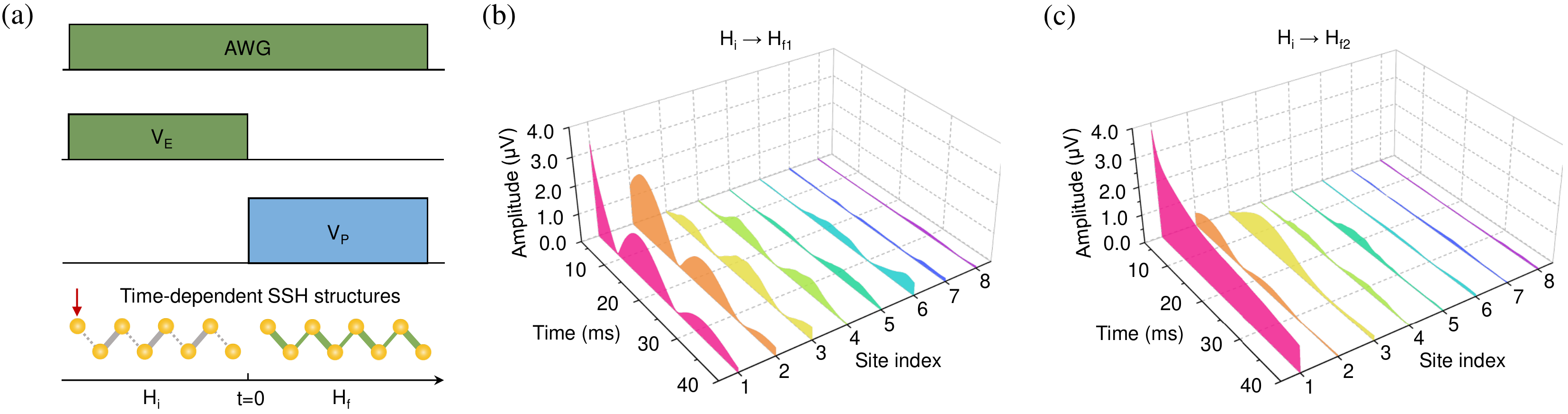}
	\caption{All amplitudes measured in the two quenches. (a) The pulse sequence used in experiments. The sinusoidal wave with the frequency $\omega_E$ generated by AWG is not only used to excite the edge oscillator (initial state preparation) in pre-quench, but also used to demodulate the edge oscillator by lock in amplifier in post-quench. The SSH structures in time evolution are realized by different coupling voltages $V_{P}$. (b) Experimental results in quench-(i). (c) Experimental results in quench-(ii). In (b) and (c), the time starts at 3 ms for the ease of display.}
	\label{pulse}
\end{figure*}

For dynamical measurements, the real-time dynamics evolution of every oscillator is demodulated by fixing its frequency. All amplitudes measured in the two quenches are showed in Fig.~\ref{pulse}, which are normalized at every moment and are showed in Fig.~\ref{RatePGP}(a) and Fig.~\ref{RatePGP}(b).

A sine wave sequence generated by a Keysight 33522B adopted as an excited source before dynamical quenching. At the same time, the sequence is the reference signal for lock-in amplifier after quenching when measuring the edge oscillator. In this way, the phase demodulated by lock-in keeps the same at every measurement and the accuracy of PGP is ensured.

\begin{table*}[t]
	\caption{\label{tab:table2} Typical voltages\footnote{All DC voltages are the same $V_{\rm DC}=4V$.} in Fig.~\ref{odd oscillators}}
	\begin{ruledtabular}
		\begin{tabular}{cccc}
			$j$  & $V_{\rm AC}^{j}$ (V) for topological&$V_{\rm AC}^{j}$ (V) for trivial  & $V_{\rm P}$     \\ \hline
			1       & 0.082          &0.250 & \multirow{2}{*}{$V_{\rm P1}=V_{\rm AC}^{1}\cos(\omega_{\rm AC}^{1}t)+V_{\rm AC}^{2}\cos(\omega_{\rm AC}^{2}t)$}             \\
			2       & 0.222          &0.075 &             \\
			3       & 0.069         &0.205 &\multirow{2}{*}{$V_{\rm P2}=V_{\rm AC}^{3}\cos(\omega_{\rm AC}^{2}t)+V_{\rm AC}^{4}\cos(\omega_{\rm AC}^{4}t)$}         \\
			4       & 0.207         &0.070 &         \\
			5       & 0.072         &0.210 & \multirow{2}{*}{$V_{\rm P3}=V_{\rm AC}^{5}\cos(\omega_{\rm AC}^{5}t)+V_{\rm AC}^{6}\cos(\omega_{\rm AC}^{6}t)$}          \\
			6       & 0.208         &0.072 &           \\
			7       & 0.072         &0.230 & $V_{\rm P4}=V_{\rm AC}^{7}\cos(\omega_{\rm AC}^{7}t)$         \\
		\end{tabular}
	\end{ruledtabular}
\end{table*}

\begin{table*}[t]
	\caption{\label{tab:table3} Typical voltages\footnote{All DC voltages are the same $V_{\rm DC}=4 V$.} in Fig.~\ref{even oscillators}}
	\begin{ruledtabular}
		\begin{tabular}{cccc}
			$j$  & $V_{\rm AC}^{j}$ (V) for topological&$V_{\rm AC}^{j}$ (V) for trivial  & $V_{\rm P}$     \\ \hline
			1    & 0.160      &0.495 & $V_{\rm P1}=V_{\rm AC}^{1}\cos(\omega_{\rm AC}^{1}t)$              \\
			2       & 0.240          &0.079 &\multirow{2}{*}{$V_{\rm P2}=V_{\rm AC}^{2}\cos(\omega_{\rm AC}^{2}t)+V_{\rm AC}^{3}\cos(\omega_{\rm AC}^{3}t)$}             \\
			3       & 0.079         &0.232 &         \\
			4       & 0.214         &0.073 &\multirow{2}{*}{$V_{\rm P3}=V_{\rm AC}^{4}\cos(\omega_{\rm AC}^{4}t)+V_{\rm AC}^{5}\cos(\omega_{\rm AC}^{5}t)$}         \\
			5       & 0.073         &0.222 &           \\
			6       & 0.220         &0.074 &\multirow{2}{*}{$V_{\rm P4}=V_{\rm AC}^{6}\cos(\omega_{\rm AC}^{6}t)+V_{\rm AC}^{7}\cos(\omega_{\rm AC}^{7}t)$}           \\
			7       & 0.085         &0.245 &          \\
		\end{tabular}
	\end{ruledtabular}
\end{table*}

\begin{table*}
	\caption{\label{table4} Parameters of the electric circuit\footnote{All Bias-Tee blocks are made up of C and L.}  in Fig.~\ref{odd oscillators} and Fig.~\ref{even oscillators}}
	\begin{ruledtabular}
		\begin{tabular}{c}
			
			R1 = 4 k$\Omega$    \\
			R = 1 M$\Omega$ \\
			C = 680 nF \\
			C1 = 1 nF \\
			L = 2.2 mH \\
			L1 = 10 mH  \\
		\end{tabular}
	\end{ruledtabular}
\end{table*}

\begin{table*}[t]
	\caption{\label{tab:disorder} Random disorders\footnote{We choose all disorders are integer because it is easily visible in split spectrum.}  $\delta J_{j}$(Hz) used in the Fig.~\ref{robutness of DPTs}}
	\begin{ruledtabular}
		\begin{tabular}{ccccccccc}
			&disorder NO.  & $j$=1  &2 & 3 & 4 & 5 & 6 & 7     \\ \hline
			\multirow{5}{*}{$\Delta$ = 5 Hz}&d1   & -5 & 1 & 1 & 5 & 3 &-2 & 0     \\
			&d2   & 0 & -5 & 4 & 2 & -3&5 & -3     \\
			&d3   & 0 & 5 & -3 &-5 & 4 &2 & 0     \\
			&d4   & -4 & -4 & -1 &-3 & -2 &-3 & -1     \\
			&d5   & -1 & 2 & -1 &-4 & -1 &1 & -2    \\	\hline	
			\multirow{5}{*}{$\Delta$ = 10 Hz}&d6   & -1 & -4 & -8 & -8 & 6 &1 & 9     \\
			&d7   & -5 & -8 & -2 & 0 & -4&4 & -1     \\
			&d8   & -8 & -2 & 9 &6 & 10 &3 & -10     \\
			&d9   & 7 & 5 & 5 &10 & -1 &6 & -2     \\
			&d10   & 3 & 9 & -10 &5 & 6 &9 & 10    \\	\hline
			\multirow{5}{*}{$\Delta$ = 15 Hz}&d11   & -10 & -4 & 4 & 9 & -13 &13 & 9     \\
			&d12   &10 & -2 & 10 & 9 & -7&-13 & -14     \\
			&d13   & -3 & -7 & 11 &14 & -4 &-13 & -10     \\
			&d14   & -15 & 5 & -11 &9 & -3 &-5 & 8     \\
			&d15   & -8 & -15 & 5 &13 & 2 &-11 & 0    \\
		\end{tabular}
	\end{ruledtabular}
\end{table*}

\end{document}